\documentclass[aps,pra,showpacs,twocolumn,superscriptaddress,longbibliography]{revtex4-1}
\usepackage{graphicx}
\usepackage[usenames]{color}
\usepackage{amssymb,amsmath}
\usepackage{siunitx}%To use degree symbol
\usepackage{xcolor}
\usepackage{setspace} 
\usepackage{float} 
\usepackage{tabularx}
\newcommand{\eq}[1]{Eq.~(\ref{#1})}
\newcommand{\fig}[1]{Fig.~\ref{#1}}
\newcommand{\be}[1]{\begin{equation}\label{#1}}
\newcommand{\ee}{\end{equation}}

\usepackage{lipsum}
\usepackage{pbox}
\usepackage[linkcolor=blue,citecolor=blue,colorlinks=true,urlcolor=blue]{hyperref}

\begin{document}

\title{ Signatures of magnetic field effects in non-sequential double ionization manifesting as back-scattering for molecules  versus forward-scattering for atoms   }

\author{G. P. Katsoulis}
\affiliation{Department of Physics and Astronomy, University College London, Gower Street, London WC1E 6BT, United Kingdom}
\author{M. B. Peters }
\affiliation{Department of Physics and Astronomy, University College London, Gower Street, London WC1E 6BT, United Kingdom}
\author{A. Staudte}
\affiliation{Joint Attosecond Laboratory, National Research Council and University of Ottawa, Ottawa, Ontario, Canada}
\author{R. Bhardwaj}
\affiliation{Department of Physics, University of Ottawa, Ottawa, ON K1N 6N5, Canada}
\author{A. Emmanouilidou}
\affiliation{Department of Physics and Astronomy, University College London, Gower Street, London WC1E 6BT, United Kingdom}

\begin{abstract}
For two-electron diatomic molecules, we investigate magnetic field effects in non-sequential double ionization where recollisions prevail. We do so by formulating a three-dimensional semi-classical model that fully accounts for the Coulomb singularities and for magnetic field effects during  time propagation. Using this model, we identify a prominent signature of non-dipole  effects. Namely, we demonstrate that   the recolliding electron  back-scatters along the direction of light propagation. Hence, this electron escapes opposite to the direction of change in momentum due to the magnetic field. This is in striking contrast to strongly-driven atoms where the recolliding electron forward-scatters along the direction of light propagation. We attribute  these distinct  signatures to the different gate that   the magnetic field creates jointly with a soft recollision in molecules compared to a hard recollision in atoms. These two different gates give rise, shortly before recollision,  to different momenta and positions of the recolliding electron 
along the direction of light propagation. As a result, we show that the Coulomb forces from the nuclei act to back-scatter the recolliding electron   in molecules and forward-scatter it in  atoms along the direction of light propagation.

 \end{abstract}
%\pacs{33.80.Rv, 34.80.Gs, 42.50.Hz}
\date{\today}

\maketitle
\section{Introduction}
 
 Strongly-driven systems constitute a computational challenge  for  ab-initio quantum-mechanical methods. 
 As a result, most theoretical studies involve different degrees of approximations and are formulated in the dipole approximation.  That is,  the space dependence of the vector potential {\bf A} is neglected. The magnetic field component of the  
Lorentz force  $\mathbf{F_B}=\mathrm{q} \mathbf{v}\times \mathbf{B}$  on a particle with charge q and velocity ${\bf v}$ is non-zero  when we account for both space and time dependence of the vector potential {\bf A}. Here, we account for magnetic field effects in  the non-relativistic regime. In this limit,   non-dipole effects are expected to arise
when the amplitude of the electron motion due to ${\bf F_B}$ is roughly equal to the Bohr radius, $\mathrm{\beta_0 \approx E_{0}^2/(8\omega c)\approx}1$ a.u. \cite{Magnetic1,Magnetic2}. Hence, for smaller frequencies $\omega$ of the field, i.e. mid-infrared wavelengths, non-dipole effects arise at   smaller  field strengths  E$_{0}$. Magnetic field effects give rise to striking features  in multi-electron ionization of systems driven by intense infrared and mid-infrared laser fields \cite{CorkumM,exp1,exp2,Ravi}.

 Non-dipole  effects have been identified  in a wide range of phenomena,  mostly,  in strongly-driven atoms \cite{Review}. For instance,  theoretical studies investigated such effects  in  multi-electron ionization of Ne  \cite{Neon},  as well as stabilization \cite{stabilization} and high-harmonic generation \cite{HHG1,HHG2,HHG3}.  In a pioneering experiment,  the average final electron momentum along the direction of light propagation was measured in single ionization of atoms \cite{CorkumM}. It  is  equal to the average kinetic energy of the electron divided by the speed of light c \cite{CorkumM}. In contrast,  this  momentum is zero in the dipole approximation.   Following this experiment,  quantum-mechanical studies      computed the electron momentum in the  direction of light propagation \cite{SI1,SI2,SI3,SI4}. They also computed  how the photon momentum I$\mathrm{_{p}}$/c   is shared between  electron and  atomic ion;  I$\mathrm{_{p}}$ is the ionization energy. This sharing  was also measured in double ionization of He by single-photon absorption \cite{exp3} and computed in strongly-driven H$_2^{+}$ with fixed nuclei \cite{SI5}.
 
 Here, we focus on  non-dipole  effects    in non-sequential double ionization (NSDI).
NSDI  involves an electron (electron 1) that tunnel-ionizes  through the field-lowered Coulomb potential. Next, this electron accelerates in the laser field
and comes back to the core to transfer  energy to the initially bound electron (electron 2), leading to the escape of both electrons. Electron-electron correlation, a fundamental interaction, underlies this field-assisted recollision   \cite{PCorkum}.  Numerous studies addressed NSDI, however, mostly in the dipole approximation  \cite{NSDI1,NSDI2}. 

\begin{figure}[ht]
\includegraphics[width=\linewidth]{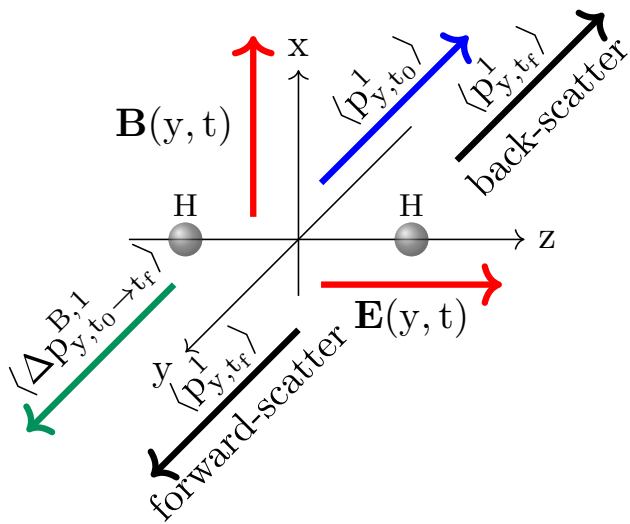}
\caption{Schematic illustration of the electric and magnetic field of the laser pulse as well as of the average initial and final momentum of the recolliding electron along the direction of light propagation.}
\label{schematic}
\end{figure}

We previously identified {\it non-dipole gated double ionization}, a prominent mechanism of non-dipole effects  in NSDI of strongly-driven atoms \cite{Emmanouilidou1,Emmanouilidou2}. The magnetic field  jointly with the recollision  act as a gate. This gate  allows for  double ionization to occur only for a  subset of the initial momenta   of the recolliding electron (electron 1) along the direction of light propagation,  +y-axis here, see \fig{schematic}.
  Namely, the average initial momentum of the recolliding electron transverse to the electric field shifts along the -y-axis. This negative shift is opposite  to the change in momentum along the +y-axis due to  ${\bf F_B}$, denoted by $\mathrm{\Delta p_{y, t_{0}\rightarrow t_f}^{B,1}}$. We predicted non-dipole gated double ionization for intensities much smaller than  expected from   $\beta_0 \approx$ 1 a.u., for strongly-driven He at 800 nm.  We found that this mechanism has a striking observable signature.  The recolliding electron forward-scatters  with a large  average final momentum along the direction of light propagation. This results  in the average sum of the final electron momenta  being roughly an order of magnitude larger than twice the average final electron momentum in single ionization. This  was  verified experimentally for driven Ar  at  2 $\mu$m \cite{FSun}. %Moreover, non-dipole gated ionization was offered as a possible explanation of double ionization yields measured in chiral molecules, i.e. molecules that are not super-imposable on their mirror image \cite{Ravi}.

 Here, we focus on the largely unexplored observable signatures of non-dipole effects  in  non-sequential double ionization  of two-electron diatomic molecules. We show that, in molecules, the recolliding electron back-scatters  with a large average final momentum opposite to the direction of light propagation, i.e. along the -y-axis.
 That is, the average electron momentum is {\it opposite} to the direction of the momentum gain due to  ${\bf F_B}$, see \fig{schematic}. This occurs mostly when the polarization of the electric field is along the molecular axis. Moreover, it takes place  for intensities significantly smaller than  expected from   $\beta_0 \approx$ 1 a.u. As a result,  the average   sum of the final electron momenta  is in the opposite direction and roughly an order of magnitude larger  than twice
  the average final momentum in single ionization.    Hence, non-dipole effects in NSDI have strikingly different signatures  in   diatomic molecules versus  atoms. The recolliding electron  back-scatters in molecules,  depending on polarization, while 
  it forward-scatters in atoms \cite{Emmanouilidou1,Emmanouilidou2}.

This difference between molecules and atoms is explained in terms of    ``soft"  \cite{Emmanouilidou3,Emmanouilidou4} versus ``hard" recollisions in non-sequential double ionization.   In a hard recollision,  electron 1 returns close to the core. It   transfers energy to  electron 2 via a long  trajectory at times roughly equal to 2T/3;   T is the period of the laser field.  In a soft recollision, electron 1 does not return as close to the core. It transfers energy to electron 2 via a short trajectory at times roughly equal to T/2.
It is well known that in a hard recollision the recolliding electron transfers energy to a bound electron through two different pathways. In the direct one, the energy transferred suffices for both electrons to ionize shortly after recollision. In the delayed pathway the energy transferred ionizes one of the two electrons shortly after recollision. The other electron transitions to an excited state and ionizes later. In contrast, in a soft recollision, electron 1 ionizes much earlier than the recollision time at the time electron 1 tunnel-ionizes through the field-lowered Coulomb potential.
Soft recollisions prevail at higher intensities  than the ones where the direct pathway  of a hard recollision prevails. The delayed pathway is prominent at small intensities \cite{Emmanouilidou3,Emmanouilidou4}. For the intensities considered here, only the direct pathway of a hard recollision is of relevance. Hence, in what follows, we refer to the direct pathway of a hard recolision as  hard recollision. The two electrons escape with a significant probability anti-parallel to each other along the electric field for a soft recollision, while they escape  parallel  to each other for  a hard one. 

We  demonstrate  signatures of  non-dipole effects in molecules for H$_2$ driven at 2000 nm.
  We find that magnetic field effects are most prominent for intensities  where soft recollisions prevail for H$_2$, while hard ones prevail for  atoms   \cite{Emmanouilidou1, Emmanouilidou2}. Here, we employ an H$_2$-like model atom   to allow   for a direct comparison with H$_2$.
 We show that the magnetic field creates a less selective gate when it acts jointly with  a soft compared to  a hard recollision.  The main differences between these two gates are as follows. Soft recollisions compared to hard ones cause a smaller negative shift 
of the average initial  momentum of the recolliding electron along the -y-axis.  Moreover, shorty before a recollision, there are three main differences: i) for a soft compared to a hard recollision,  the initial  momentum maps  to much smaller values of  positive momenta of  electron 1; ii) electron 1 has mostly  small positive  positions for a soft recollision versus  mostly large negative positions for a hard one; iii) due to differences i) and ii),  the Coulomb attractive forces  from the nuclei  back-scatter  electron 1 for a soft recollision while they forward-scatter it for a hard one.
 Back-scattering of  the recolliding electron  occurs for a wide range of wavelengths and intensities. For  these parameters, $\beta_0$ is significantly smaller than one  where magnetic fields are expected to arise.

%In hard (soft) recollisions,  electron 1 returns close (not as close) to the core. It   transfers energy to  electron 2 via a long (short)  trajectory at times roughly equal to 2T/3 (T/2);   T is the period of the laser field. Recollision-assisted field ionization (soft) occurs at higher intensities compared to   field-assisted recollision (hard).  The two electrons escape mostly parallel to each other  along the  direction of polarization of the electric field for hard recollisions versus mostly anti-parallel for soft ones  \cite{Emmanouilidou3,Emmanouilidou4}.
% Here, we show that the magnetic field creates a less selective gate when it acts jointly with  a soft compared to a hard recollision. Soft recollisions cause a smaller negative shift 
%of the average initial transverse momentum of electron 1. A small (large) initial negative shift maps shortly before a soft (hard) recollision to mostly small (large) positive transverse momenta of electron 1. Moreover, it maps shortly before a soft (hard) recollision to mostly  small positive (large negative) position of electron 1 along the propagation direction. This results in the Coulomb attractive forces  from the nuclei acting to back-scatter electron 1 for a soft  and to forward-scatter it  for a hard recollision.

   %The latter is driven by the same pulse and shares the  same total nuclear charge, two, and the same first and second ionization potentials as H$_2$. 
 In the current study of strongly-driven H$_2$,  we keep the nuclei fixed. The reason is     that NSDI prevails double ionization of H$_2$ for fixed   but not for Coulomb exploding nuclei.  As a result,  non-dipole gated double ionization prevails double ionization mostly for strongly-driven H$_2$ with fixed nuclei. 
  Our computations can serve as benchmark  for molecules which doubly ionize via recollisions and Coulomb explode much slower than  the duration of the laser pulse.
 Re-enforcing the applicability of our findings for  fixed nuclei, we also identify signatures of electron 1  back-scattering   in differential observables of double ionization during the break-up of driven H$_2$.

 Finally, we formulate a three-dimensional semi-classical technique that accounts for both  Coulomb singularities and non-dipole effects during time propagation of driven two electron diatomic molecules. Employing this technique allows us to demonstrate the above mentioned magnetic field effects in  molecules. This model employs a leapfrog technique for time propagation of molecules that is more involved than the one
previously employed in the dipole approximation \cite{toolkit2014}. The higher complexity is due to the derivatives of the momenta depending on the momenta and the derivatives of the positions depending on the positions of the particles in the equations of motion.

 \section{Method}
Here, we formulate a three-dimensional (3D) semi-classical model for strongly-driven two-electron diatomic molecules that fully accounts for magnetic field effects during time propagation. The backbone of the current model is a 3D semi-classical model   for strongly-driven diatomic molecules developed in the dipole approximation
\cite{toolkit2014}. Successes of this latter model include  identifying   the prevalent pathways of Rydberg formation during the break-up of strongly-driven H$_2$ \cite{RydbergAE1} and D$_{3}^{+}$ \cite{RydbergAE2}. Our results for Rydberg formation  were   in very good agreement with experimental results \cite{ExpR1,ExpR2}. There is a major difference between the current  and the previous 3D semi-classical model. Here, in the equations of motion, due to the force $\mathrm{\bf F_B}$, the derivatives of the positions depend on the positions and the derivatives of the momenta depend on the momenta of the particles. Hence,  to   integrate  the equations of motion that include non-dipole effects, we adopt a leapfrog technique  \citep{Pihajoki2015,Liu2016}  more involved than the  one used in the dipole approximation  \cite{toolkit2014}. 

 We use the  initial state  employed in the 3D semi-classical model in the dipole approximation.   The time t$_0$ when the time propagation starts is selected randomly within the time interval of non-zero values of the electric field. The nuclei are initiated at rest, since we previously found that an initial predissociation does not significantly alter the ionization dynamics \cite{toolkit2014}. The initially bound electron 2 is described by  a microcanonical distribution \cite{toolkit2014}.  If the 
 electric field strength  at time t$_0$  is within the below-the-barrier ionization regime,  electron 1 tunnel-ionizes through the field-lowered Coulomb potential.  This is also the time   the propagation starts. We employ a non-relativistic quantum-mechanical formula to compute the tunneling rate \cite{tun}. We assume that electron 1 tunnels along the direction of the electric field with zero momentum.   The  momentum transverse to the electric field is described by a Gaussian distribution that  represents the Gaussian-shaped filter with an intensity-dependent width arising from standard tunnelling theory \cite{tun1,tun2,tun3}. In the dipole approximation,  the Gaussian distribution is centered around zero.  
  If the field strength   is within the over-the-barrier ionization regime,  electron 1  tunnel-ionizes at the maximum of the field-lowered Coulomb potential.  We set the kinetic energy of electron 1 equal to the difference between the first ionization energy of H$_2$ and the maximum of the field-lowered Coulomb potential \cite{toolkit2014}. The polar and azimuthal angles of the  momentum of electron 1 are selected randomly with the restriction that this momentum is opposite to the direction of the electric field \cite{toolkit2014}. Hence, the initial transverse electron momentum  is centered around zero. 

 During time propagation of  strongly-driven H$_2$, each electron is allowed to tunnel  if the   nuclei are moving, while tunneling is switched off   if the nuclei are fixed. The reason we do so is that NSDI prevails double ionization for driven H$_2$ with fixed nuclei. However,
  NSDI competes with enhanced ionization \cite{Niikura,EI1,EI2,EI3,EI4} during the break-up of strongly-driven H$_2$.     To account for enhanced ionization, we need to allow each electron to tunnel during time propagation.  Enhanced ionization   occurs at a critical distance of the nuclei. A double-potential well is formed such that it is easier for an electron bound to the higher potential well to tunnel to the lower one  and then  ionize. For moving nuclei, we allow for each electron to tunnel 
  with a non-relativistic quantum-mechanical probability given by the Wentzel-Kramers-Brillouin approximation \cite{toolkit2014}. 
 Moreover, in our computations of the break-up of strongly-driven H$_2$, the motion of the electrons and the nuclei are treated on an equal footing.  
 In contrast to NSDI, enhanced ionization does not require electron-electron correlation. As a result, non-dipole gated double ionization does not occur in enhanced ionization.

  In this work, non-dipole effects are not included in tunneling that occurs in the initial state if the field strength is below the barrier. 
  In Ref. \cite{keitel}, magnetic field effects were accounted for in the initial tunneling rate of an atom via a relativistic formulation.
 As a result, at the exit point, the  momentum of electron 1 along the direction of light propagation  is centered around a non-zero momentum.  For the intensities considered in this work, the largest value of this  momentum is  0.04 $\mathrm{I_p/c}$ \cite{keitel}. For H$_{2}$, with $\mathrm{I_p}$ equal to 0.567 a.u., 
  this momentum is equal to
 %This momentum ranges  from 0.33 $\mathrm{I_p/c}$ to almost zero with increasing $\mathrm{E_0/(2I_p)^{3/2}}$. Here, the smallest intensity considered is  2$\times$10$^{14}$ W/cm$^{2}$. For this intensity,  for H$_2$ with $\mathrm{I_p}$ equal to 0.567 a.u., the most probable initial transverse electron momentum  is 0.04 $\mathrm{I_p/c}$ \cite{keitel}, i.e. 
 1.7$\times$10$^{-4}$ a.u. 
   This value is much smaller compared to all average final electron momenta along the y-axis obtained in this work. Moreover, for the results presented here, for the vast majority  (at least 80\%) of double ionization events, the field strength is over the barrier. For these events, there is no extra shift due to non-dipole effects in the initial state \cite{keitel}. Hence,  we safely ignore this initial shift for all double ionization events \cite{Emmanouilidou1,Emmanouilidou2}.   
% To  obtain our results for the H$_2$-like atom, we employ a 3D semi-classical technique for strongly-driven atoms that  fully accounts for non-dipole effects during time propagation of H$_2$. This  technique does not involve tunneling during time propagation. It was used to identify non-dipole gated double ionization in strongly-driven He and Xe  \cite{Emmanouilidou1,Emmanouilidou2}.  As for molecules, we do not account for non-dipole effects in the tunneling   in the initial state. 

For molecules, we employ the Hamiltonian of the strongly driven four-body system to propagate classically the positions and the momenta of all particles. All Coulomb forces and the interaction of each electron and nucleus with the laser field  are fully accounted for with no approximation. The Hamiltonian  is given by
\begin{equation}\label{Hamiltonian}
\mathrm{H} = \sum_{\mathrm{i}=1}^{\mathrm{N}}\frac{\left(\mathbf{\tilde{p}}_{\mathrm{i}}- \mathrm{Q_i}\mathbf{A}(\mathbf{r}_{\mathrm{i}},\mathrm{t}) \right)^2}{2\mathrm{m_i}}+\sum_{\mathrm{i}=1}^{\mathrm{N-1}}\sum_{\mathrm{j}=\mathrm{i}+1}^{\mathrm{N}}\frac{\mathrm{Q_i}\mathrm{Q_j}}{|\mathbf{r_i}-\mathbf{r_j}|},
\end{equation}
where $\mathrm{N=4}$,  while $\mathrm{Q_i}$ is the charge, $\mathrm{m_i}$ is the mass, $\mathbf{r}_{\mathrm{i}}$ is the position vector and $\mathbf{\tilde{p}}_{\mathrm{i}}$ is the canonical momentum vector of particle i. The mechanical momentum is given by $\mathrm{\mathbf{p}_{i}=\mathbf{\tilde{p}}_{\mathrm{i}}- \mathrm{Q_i}\mathbf{A}(\mathbf{r}_{\mathrm{i}},\mathrm{t}})$. The results presented in section \ref{results} concern the mechanical momentum of the electrons. The vector potential  we employ is given by
\begin{equation}\label{eq:vector_potential}
\mathbf{A}(\mathrm{y,t}) = -\frac{\mathrm{E}_0}{\omega}\exp \left[ - 2\ln (2)\left( \frac{\mathrm{c t - y}}{\mathrm{c \tau}} \right)^2 \right]   \sin ( \omega \mathrm{t}  - \mathrm{k y}+\mathrm{\phi}) \hat{\mathbf{z}},
\end{equation}
where $\mathrm{k=\omega/c}$ is the  wave number of the
electric field and  $\mathrm{\tau}$ is the full width at half maximum of the pulse duration in intensity. The carrier envelope phase (CEP) is denoted by $\phi$. This phase is important only for small duration laser pulses. Hence, for our computations at 2000 nm, $\phi$ is zero when we consider a  40 fs duration pulse, while it is non zero when we consider a near-single-cycle pulse of 10 fs duration in section \ref{results}.  The  electric field  is taken to be along the z-axis and the propagation direction  is along the +y-axis. The molecular axis is changing depending on the direction of polarization of the electric field with respect to the molecular axis.

\subsection{Propagation Technique}
 In our classical formulation, we fully account for the Coulomb singularities. Hence, an electron can approach infinitely close to the nucleus during time propagation. To ensure the accurate numerical treatment of the N-body problem in the laser field, we perform a global regularisation.  This regularisation was introduced in the context of the gravitational N-body problem  \cite{Heggie1974}. Here, we integrate the equations of motion using  a leapfrog technique  \citep{Pihajoki2015,Liu2016} jointly with  the Bulirsch-Stoer method \cite{press2007numerical,bulirsch1966numerical}. This ensures  numerical accuracy.   Also, we employ  a more involved leapfrog technique than the 
 one previously employed  in our  studies of strongly-driven two-electron  molecules in the dipole approximation  \cite{toolkit2014}. We do so, since, here, the  derivatives of the positions and the momenta   depend   on the quantities themselves.

\subsubsection{Equations of motion}
 The Hamiltonian of the N-body problem  is defined in \eq{Hamiltonian}. Next,  we transform to a new   coordinate system that  involves the relative positions $\mathbf{q}$ and the corresponding conjugate momenta $\boldsymbol{\rho}$ of the particles. We define
 \begin{equation}
\mathbf{q}_{\mathrm{ij}}=\mathbf{r}_{\mathrm{i}}-\mathbf{r}_{\mathrm{j}}
\end{equation}
and
\begin{equation}
\boldsymbol{\rho}_{\mathrm{ij}}=\frac{1}{\mathrm{N}}\left( \mathbf{\tilde{p}}_{\mathrm{i}} - \mathbf{\tilde{p}}_{\mathrm{j}}- \frac{\mathrm{m_i}-\mathrm{m_j}}{\mathrm{M}}\langle \boldsymbol{\rho} \rangle \right) ,
\end{equation}
where
\begin{equation}
\langle \boldsymbol{\rho} \rangle = \sum_{\mathrm{i=1}}^{\mathrm{N}}\mathbf{\tilde{p}}_{\mathrm{i}} \; \; \text{and} \; \; \mathrm{M}=\sum_{\mathrm{i=1}}^{\mathrm{N}}\mathrm{m_i}.
\end{equation}
%and $\mathrm{M}$ the sum of the masses i.e.
%\begin{equation}
%\mathrm{M}=\sum_{\mathrm{i=1}}^{\mathrm{N}}\mathrm{m_i}.
%\end{equation}
The inverse transformation is given by
\begin{equation}\label{position_in_terms_of_q}
\mathbf{r}_\mathrm{i}=\frac{1}{\mathrm{M}}\sum_{\mathrm{j=i+1}}^{\mathrm{N}}\mathrm{m_j}\mathbf{{q}}_{\mathrm{ij}}-\frac{1}{\mathrm{M}}\sum_{\mathrm{j=1}}^{\mathrm{i-1}}\mathrm{m_j}\mathbf{q}_{\mathrm{ji}}+ \langle \mathbf{q} \rangle,
\end{equation}
and
\begin{equation}\label{momenta_in_terms_of_rho}
\mathbf{\tilde{p}}_{\mathrm{i}}=\sum_{\mathrm{j=i+1}}^{\mathrm{N}}\boldsymbol{\rho}_{\mathrm{ij}}-\sum_{\mathrm{j=1}}^{\mathrm{i-1}}\boldsymbol{\rho}_{\mathrm{ji}}+ \frac{\mathrm{m_i}}{\mathrm{M}} \langle \boldsymbol{\rho} \rangle,
\end{equation}
where
\begin{equation}
\langle \mathbf{q} \rangle = \frac{1}{\mathrm{M}} \sum_{\mathrm{i=1}}^{\mathrm{N}}\mathrm{m_i}\mathbf{r}_{\mathrm{i}}.
\end{equation}
One can simplify Eqs. \eqref{position_in_terms_of_q} and \eqref{momenta_in_terms_of_rho} by defining a fictitious particle $\mathrm{k},$ corresponding to each pair of particles $\mathrm{ij}.$ We define the index $\mathrm{k}$ as
\begin{equation}
\mathrm{k(i,j)} = \mathrm{ (i-1)N - i(i+1)/2 + j },
\end{equation}
with $\mathrm{j>i}$ and the  total number of fictitious particles being $\mathrm{K=N(N-1)/2}$.
In addition, we define the parameters $\alpha_{\mathrm{ik}}$ and $\beta_{\mathrm{ik}},$ as $\alpha_{\mathrm{ik}}=1,\beta_{\mathrm{ik}}=\mathrm{m_j/M}$ and $\alpha_{\mathrm{jk}}=-1,\beta_{\mathrm{jk}}=-\mathrm{m_i/M}$ when $\mathrm{k=k(ij)}$, otherwise, $\alpha_{\mathrm{ik}}=\beta_{\mathrm{ik}}=0$. %We present the values of these  parameters  in Table \ref{tab:alpha_beta}.
Eqs. \eqref{position_in_terms_of_q} and \eqref{momenta_in_terms_of_rho} now take the form 
\begin{equation}
\mathbf{\tilde{p}}_{\mathrm{i}}=\sum_{\mathrm{k=1}}^{K}\alpha_{ik}\boldsymbol{\rho}_{\mathrm{k}}+ \frac{\mathrm{m_i}}{\mathrm{M}} \langle \boldsymbol{\rho} \rangle,
\end{equation}
and
\begin{equation}
\mathbf{r}_{\mathrm{i}}=\sum_{\mathrm{k=1}}^{K}\beta_{ik}\mathbf{q}_{\mathrm{k}}+ \langle \mathbf{q} \rangle.
\end{equation}
The Hamiltonian of the system can now be written in terms of the relative positions  $\mathbf{q}$ and the corresponding conjugate momenta $\boldsymbol{\rho}$ as

%\begin{widetext}\label{Hamiltonian_in_transformed_coordinates}
%\begin{align}
%\begin{split}
%\mathrm{H} &= \sum_{\mathrm{k,k'=1}}^{\mathrm{K}}\mathrm{T}_{\mathrm{kk'}}\boldsymbol{\rho}_{\mathrm{k}}\boldsymbol{\rho}_{\mathrm{k'}}+\frac{\langle \boldsymbol{\rho}\rangle^2}{2M}+ \sum_{\mathrm{i=1}}^{\mathrm{N}}\frac{\mathrm{Q^2_i}}{2\mathrm{m_i}}\left[ \mathcal{\mathbf{A}}\left( \sum_{\mathrm{k=1}}^{\mathrm{K}}\beta_{ik}\mathbf{q}_{\mathrm{k}} + \langle \mathbf{q} \rangle ,\mathrm{t} \right) \right]^2 \\
%&\hspace{5cm}- \sum_{\mathrm{i=1}}^{\mathrm{N}}\frac{\mathrm{Q_i}}{\mathrm{m_i}}\left[ \sum_{\mathrm{k=1}}^{\mathrm{K}}\alpha_{ik}\boldsymbol{\rho}_{\mathrm{k}} + \frac{\mathrm{m_i}}{\mathrm{M}} \langle \boldsymbol{\rho} \rangle \right] \cdot \left[ \mathbf{A}\left( \sum_{\mathrm{k=1}}^{\mathrm{K}}\beta_\mathrm{ik}\mathbf{q}_{\mathrm{k}} + \langle \mathbf{q} \rangle ,\mathrm{t} \right) \right]+\sum_{\mathrm{k=1}}^{\mathrm{K}}\frac{\mathrm{U_k}}{\mathrm{q_k}}
%\end{split}
%\end{align}
%\end{widetext}

\begin{align}\label{Hamiltonian_in_transformed_coordinates}
\begin{split}
\mathrm{H} &= \sum_{\mathrm{k,k'=1}}^{\mathrm{K}}\mathrm{T}_{\mathrm{kk'}}\boldsymbol{\rho}_{\mathrm{k}}\boldsymbol{\rho}_{\mathrm{k'}}+\frac{\langle \boldsymbol{\rho}\rangle^2}{2M}+\sum_{\mathrm{k=1}}^{\mathrm{K}}\frac{\mathrm{U_k}}{\mathrm{q_k}}\\
& +\sum_{\mathrm{i=1}}^{\mathrm{N}}\frac{\mathrm{Q^2_i}}{2\mathrm{m_i}}\mathbf{A}^2\left( \mathbf{r}_{\mathrm{i}} ,\mathrm{t} \right)  - \sum_{\mathrm{i=1}}^{\mathrm{N}}\frac{\mathrm{Q_i}}{\mathrm{m_i}}\mathbf{\tilde{p}}_{\mathrm{i}} \cdot \mathbf{A}\left( \mathbf{r}_{\mathrm{i}} ,\mathrm{t} \right), 
\end{split}
\end{align}

with 
\begin{equation}
\mathrm{T_{kk'}}=\sum_{i=1}^{\mathrm{N}}\dfrac{\alpha_{\mathrm{ik}}\alpha_{\mathrm{ik'}}}{2\mathrm{m_i}} \; \; \text{and} \; \; \mathrm{U_k} = \mathrm{Q_iQ_j}.
\end{equation}

The equations of motion are  expressed as follows
\begin{align}\label{eq:Equations_of_motion}
\begin{split}
\frac{\mathrm{d}\mathbf{q}_{\mathrm{k}}}{\mathrm{dt}}&=2\sum_{\mathrm{k'}=1}^{\mathrm{K}}\mathrm{T}_{\mathrm{kk'}}\boldsymbol{\rho}_{\mathrm{k'}}-\sum_{\mathrm{i=1}}^{\mathrm{N}}\frac{\mathrm{Q_i}}{\mathrm{m_i}} \alpha_{\mathrm{ik}}\mathbf{A}\left( \mathbf{r}_{\mathrm{i}} ,\mathrm{t} \right) \\
\frac{\mathrm{d}\langle \mathbf{q} \rangle}{\mathrm{dt}}&=\dfrac{1}{\mathrm{M}}\langle \boldsymbol{\rho} \rangle\\
\frac{\mathrm{d}\boldsymbol{\rho}_{\mathrm{k}}}{\mathrm{dt}}&= \frac{\mathrm{U_k}\mathbf{q}_{\mathrm{k}}}{\mathrm{q^3_k}} + \sum_{\mathrm{i=1}}^{\mathrm{N}}\frac{\mathrm{Q_i}}{\mathrm{m_i}} \left( \mathbf{\tilde{p}}_{\mathrm{i}} - \mathrm{Q_i}\mathbf{A}\left( {\mathbf{r}}_{\mathrm{i}},\mathrm{t} \right)\right)\cdot\dfrac{\partial\mathbf{A}\left( {\mathbf{r}}_{\mathrm{i}},\mathrm{t} \right)}{\partial \mathbf{q_k}} \\
\frac{\mathrm{d}\langle \boldsymbol{\rho}\rangle}{\mathrm{dt}}&=\sum_{\mathrm{i=1}}^{\mathrm{N}}\frac{\mathrm{Q_i}}{\mathrm{m_i}} \left( \mathbf{\tilde{p}}_{\mathrm{i}} - \mathrm{Q_i}\mathbf{A}\left( {\mathbf{r}}_{\mathrm{i}},\mathrm{t} \right)\right)\cdot\dfrac{\partial\mathbf{A}\left( {\mathbf{r}}_{\mathrm{i}},\mathrm{t} \right)}{\partial \langle \mathbf{q} \rangle}.
\end{split}
\end{align}
 
\subsubsection{Time-transformed leapfrog technique } 
In the equations of motion \eqref{eq:Equations_of_motion}  the derivatives of the positions depend on the positions and the derivatives of the momenta depend on the momenta of the particles.  To integrate  these equations, we employ a leapfrog technique  \citep{Pihajoki2015,Liu2016}  outlined in what follows. First,  we perform a time transformation $\mathrm{t}\to\mathrm{s}$, where
\begin{equation}
\mathrm{ds}=\Omega(\mathbf{q})\mathrm{dt},
\end{equation}
with $\Omega(\mathbf{q})$ an arbitrary positive function of $\mathbf{q}.$ We select the function
\begin{equation}
\Omega(\mathbf{q}) = \sum_{\mathrm{k=1}}^{\mathrm{K}}\dfrac{1}{|\mathbf{q}_\mathrm{k}|},
\end{equation}
which forces  the time step to decrease when two particles undergo a close encounter and to increase when all particles are far away from each other. The equations of motion now take the following form
\begin{align}
\begin{split}
\mathbf{q'}&=\dot{\mathbf{q}}(\mathbf{q},\boldsymbol{\rho},\mathrm{t})/\Omega(\mathbf{q})\\
\boldsymbol{\rho'}&=\dot{\boldsymbol{\rho}}(\mathbf{q},\boldsymbol{\rho},\mathrm{t})/\Omega(\mathbf{q})\\
\mathrm{t'}&=1/\Omega(\mathbf{q}),
\end{split}
\end{align}
with prime denoting the derivative with respect to the new variable s. To perform the integration  we use the leapfrog technique described in \citep{Pihajoki2015,Liu2016}. This leapfrog technique introduces three auxiliary variables, two vectors $\mathbf{W^q},\mathbf{W}^{\boldsymbol{\rho}}$ and one scalar $\mathrm{W^t}$. 
As a result, an extended system is obtained where the derivatives of the position and the momenta no longer depend on the quantities themselves.
 The   extended equations are given by
\begin{align*}
\begin{split}
\mathbf{q'}&=\dot{\mathbf{q}}(\mathbf{W^q},\boldsymbol{\rho},\mathrm{W^t})/\Omega(\mathbf{W^q})\\
\mathbf{W}^{\boldsymbol{\rho}'}&=\dot{\boldsymbol{\rho}}(\mathbf{W^q},\boldsymbol{\rho},\mathrm{W^t})/\Omega(\mathbf{W^q})\\
\mathrm{t'}&=1/\Omega(\mathbf{W^q}),
\end{split}
\end{align*}
and
\begin{align*}
\begin{split}
\mathbf{W^{q'}}&=\dot{\mathbf{q}}(\mathbf{q},\mathbf{W^{\boldsymbol{\rho}}},\mathrm{t})/\Omega(\mathbf{q})\\
\boldsymbol{\rho'}&=\dot{\boldsymbol{\rho}}(\mathbf{q},\mathbf{W^{\boldsymbol{\rho}}},\mathrm{t})/\Omega(\mathbf{q})\\
\mathrm{W^{t'}}&=1/\Omega(\mathbf{q}).
\end{split}
\end{align*}

\begin{figure}[ht]
\includegraphics[width=\linewidth]{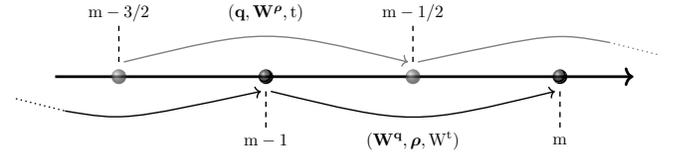}
\caption{Schematic illustration of the propagation of the two triplets $(\mathbf{q},\mathbf{W}^{\boldsymbol{\rho}},\mathrm{t})$ and $(\mathbf{W^q},\boldsymbol{\rho},\mathrm{W^t})$ over a substep of size h, $\mathrm{m-3/2}\to\mathrm{m-1/2}$ and $\mathrm{m-1}\to\mathrm{m}$ respectively, with m=2,...,n. }\label{fig:leapfrog_illustration}
\end{figure}

We propagate for a  time step, by propagating for half a step each triplet of variables ($\mathbf{q},\mathbf{W}^{\boldsymbol{\rho}}$,t) and ($\mathbf{W^q},\boldsymbol{\rho},\mathrm{W^t}$)  in an alternating way, see the leapfrog algorithm described in the  Appendix\ref{Appendix}. 
Moreover, to achieve better accuracy, we incorporate the leapfrog method in the Bulirsch-Stoer extrapolation scheme  \cite{press2007numerical,bulirsch1966numerical}.   In this scheme, a propagation over a step H, is split into n sub steps of size $\mathrm{h=H/n}.$ We use the leapfrog method to propagate over each substep. In \fig{fig:leapfrog_illustration}, we offer a schematic illustration of the propagation during a time substep of size h. The detailed algorithm is described in the Appendix\ref{Appendix}.  This process  is repeated with increasing number of substeps, i.e. n$\to \infty,$ until an extrapolation with a satisfactory error is achieved. 

 \subsubsection{Tunnelling during time propagation of H$_2$ with moving nuclei}
During the break-up of strongly-driven  H$_2$, we allow for each electron to tunnel at the classical turning points along the electric field axis using the Wentzel-Kramers-Brillouin (WKB) approximation \cite{toolkit2014}. Each electron i can tunnel through a potential barrier given by 
\begin{equation}
\mathbf{r}_{\mathrm{i}}\cdot \mathbf{E}(\mathbf{r}_{\mathrm{i}},\mathrm{t})+\mathrm{V}(\mathbf{r}_\mathrm{i}),
\end{equation}
where $\mathbf{E}(\mathbf{r}_{\mathrm{i}},\mathrm{t})$ is the electric field and $\mathrm{V}(\mathbf{r}_\mathrm{i})$ is the Coulomb interaction with the nuclei. The term in the barrier involving the electric field does not appear in the Hamiltonian \eqref{Hamiltonian}. There is no contradiction.  Indeed, the gauge-invariant energy of a particle does not always coincide with the gauge-dependent 
hamiltonian, as discussed in Ref. \cite{kobe1987,keitel}. The energy of electron i is given by 
\begin{equation}
\dfrac{(\mathbf{\tilde{p}}_{\mathrm{i}}+\mathbf{A}(\mathbf{r}_{\mathrm{i}},\mathrm{t}))^2}{2}+\mathbf{r}_{\mathrm{i}}\cdot \mathbf{E}(\mathbf{r}_{\mathrm{i}},\mathrm{t})+\mathrm{V}(\mathbf{r}_\mathrm{i}).
\end{equation}
% We note once more that  we ignore the magnetic field effects during tunneling in the time propagation of H$_2$ with moving nuclei.  
%
%%%%%%%%%%%%%%%%%%%%%%%%%%%%%%%%%%%%%%%%%%%%%%%%%%%%%%%%%%
%%%%%%%%%%%%      RESULTS
%%%%%%%%%%%%%%%%%%%%%%%%%%%%%%%%%%%%%%%%%%%%%%%%%%%%%%%%%%
%\newpage

\section{Results}\label{results}
%\lipsum
For our computations,  the propagation of H$_2$ or of the H$_2$-like atom starts at time $\mathrm{t_0}$ and ends at time t$_\mathrm{f}$ when the energies of the two electrons converge. If the energies of both electrons are positive, these events correspond to double ionization (DI). Moreover, if during time propagation a recollision takes place, 
we label these events as NSDI. To identify a recollision, we monitor the inter-electronic repulsive potential energy during  time propagation. The time when this potential energy becomes maximum corresponds to the two electrons having a minimum approach. We register this time as the time of recollision, t$\mathrm{_{r}}$. Also, we define the ionization time, $\mathrm{t_{i}^{ion}}$, 
of an electron i  as the time when the compensated energy $\mathrm{(p_{x,i}^{2}+p_{y,i}^2+(p_{z,i}-\hat{\mathbf{z}}\cdot\mathbf{A}(y_{i},t))^2)/2+V(\mathbf{r}_{i})}$ becomes positive and remains positive thereafter  \cite{Leopold}, as in previous studies \cite{Emmanouilidou4}.
For the intensities and wavelengths considered here, the vast majority of non-sequential double ionization events  have only one recollision. Hence,  we only consider NSDI events with one recollision. Moreover, for the pulse parameters considered here, for fixed nuclei, the majority of double ionization events are NSDI events, unless we mention otherwise.  In our computations, the uncertainty of the average electron momentum is proportional to $1/\sqrt{\mathrm{n_{ev}}}$, with n$_{\mathrm{ev}}$ being the number of events for single or double ionization.  For all pulse parameters,  we obtain at least 1.5$\times$10$^{5}$ number of DI events, rendering  the uncertainty  very small.

\subsection{Strongly-driven H$_2$ with fixed nuclei}

We focus on non-dipole effects in  H$_2$ with fixed nuclei driven by a laser pulse of intensity 3.5$\times$10$^{14}$ W/cm$^2$ and 40 fs duration at 2000 nm.  %Our results  serve as a benchmark  for molecules Coulomb exploding in a time scale larger  than  the duration of the pulse.  In these cases, double ionization is mostly due to NSDI.
The uncertainty in our computations is roughly 4$\times$10$^{-3}$ a.u.
\subsubsection{Single ionization}

For single ionization, we compute the average of the final electron momentum 
%%%%%%%%%%%%%%%%%%%% Table for single ionizaiton
\begin{table}[ht]
\caption{\label{tab:si_average_mom} Average final electron momentum along the direction of light propagation   and average kinetic energy  of  $\mathrm{H_2}$  with parallel polarization and fixed nuclei  in single ionization.  The Coulomb forces are either switched  off or on. The values are expressed  in  $10^{-3}$ a.u. } 
\begin{ruledtabular}
\begin{tabular}{lclclcl}
& \multicolumn{3}{c} { \hspace{-1.2cm} Single ionization of  $\mathrm{H_2}$ $\parallel$ with fixed nuclei}  \\
\hline
 & \multicolumn{3}{c}{ \textrm{Coulomb off} \;\;\;\;\;\;\; \textrm{Coulomb on} }\\
\hline
$\langle \mathrm{p^1_{y,t_{f}}}\rangle\mathrm{_{SI}}$ & \hspace{1.5cm}  6 & \hspace{2cm} 4  \\
$\langle \mathrm{E^1_{k}/c}\rangle\mathrm{_{SI}}$ &  \hspace{1.5cm}  6 &  \hspace{2cm} 8  \\
\end{tabular}
\end{ruledtabular}
\end{table}
$\mathrm{\langle p_{y,t_{f}}^1\rangle_{SI}}$, along the direction of light propagation, i.e.  along the +y-axis, see Table \ref{tab:si_average_mom}. 
First, we switch off all   Coulomb forces. We find that  the average final  momentum of electron 1 is equal to its average kinetic energy, $\mathrm{\langle E_k\rangle}$, divided by the speed of light c, as expected \cite{CorkumM,SI5}. For Coulomb forces  switched on, the average final electron momentum along the y-axis is still positive. However, it is equal to  4$\times$10$^{-3}$ a.u. and  is no longer given by $\mathrm{\langle E_k\rangle/c}$, which is  equal to 8$\times$10$^{-3}$ a.u.

\subsubsection{Average electron momenta in double ionization}
Next, we focus on non-sequential double ionization of H$_2$ with fixed nuclei driven by a pulse of either parallel or perpendicular polarization with respect to the molecular axis. In Table \ref{tab:average_mom},  we compare the results for H$_2$  with the results for   the H$_2$-like atom for NSDI  and for all double ionization events. The H$_2$-like atom  is driven by the same pulse and shares the  same total nuclear charge, two, and the same first and second ionization potentials as H$_2$.
 For all three cases, in non-sequential double ionization, the average of the initial momentum 
of electron 1 along the direction of light propagation  has a negative shift. That is, $\mathrm{\langle p^1_{y,t_0}\rangle_{NSDI}}$  is shifted opposite to the direction of the momentum change due to the magnetic  force  ${\bf F_B}$. However, this initial negative shift is much larger for the H$_2$-like atom
compared to H$_2$. Indeed, this shift for the H$_2$-like atom  has more than double (three times) the value for   H$_2$ with parallel (perpendicular) polarization.  For simplicity, we refer to the momentum and position of an electron along the direction of light propagation as transverse. Moreover, for  H$_2$ with parallel polarization, the average final transverse   momentum of electron 1, i.e. $\mathrm{\langle p_{y,t_{f}}^1\rangle_{NSDI}}$, and the average of the sum of the final transverse electron momenta, i.e. $\mathrm{\langle p_{y,t_{f}}^1+p_{y,t_{f}}^2\rangle_{NSDI}}$, are almost an order of magnitude larger than twice the average final electron momentum for single ionization, i.e. 2$\mathrm{\langle p_{y,t_{f}}^1\rangle_{SI}}$. Similarly, we find that $\mathrm{\langle p_{y,t_{f}}^1\rangle_{NSDI}}$ is more than a magnitude larger than 2$\mathrm{\langle p_{y,t_{f}}^1\rangle_{SI}}$ for the H$_2$-like atom. However, there is a  striking difference between H$_2$ with parallel polarization and the H$_2$-like atom. In the former case, $\mathrm{\langle p_{y,t_{f}}^1\rangle_{NSDI}}$ and  $\mathrm{\langle p_{y,t_{f}}^1+ p_{y,t_f}^2\rangle_{NSDI}}$ have large negative values, while in the latter  case they have   large positive values. Hence,  on average, electron 1  back-scatters for  H$_2$ with parallel polarization and forward-scatters for the H$_2$-like atom along the direction of light propagation. 
 Forward-scattering of electron 1  is in accord with our previous finding for strongly-driven He at 800 nm \cite{Emmanouilidou1,Emmanouilidou2}.  Also, we find that for H$_{2}$ with perpendicular polarization the average final electron momentum has small values. 

 We also compute NSDI for H$_{2}$ with fixed nuclei when driven by a near-single-cycle laser pulse of parallel polarisation, intensity of 3.5 $\times 10^{14}$ W/cm$^2$ and 10 fs duration at 2000 nm. We consider five values of  $\phi$ in \eq{eq:vector_potential}, i.e. of CEP, from 0$^{\circ}$ to 360$^{\circ}$ in steps of 60$^{\circ}$. Such CEP-controlled near-single-cycle  laser pulses have been employed in experiments of NSDI \cite{singlec1}. 
Averaging over all CEPs, we find that $\langle \mathrm{p^1_{y,t_{f}}} + \mathrm{p^2_{y,t_{f}}}\rangle_{\mathrm{NSDI}}=-110\times 10^{-3}$ a.u. That is, when  H$_2$ is driven by a 10 fs duration pulse we find that the recolliding electron back-scatters as is the case for a 40 fs duration pulse, see Table \eqref{tab:average_mom}. This corroborates our previous statement that our results for H$_2$ with fixed nuclei driven by a 40 fs duration pulse apply to molecules  Coulomb exploding in a timescale larger than the duration of the pulse. Computations with a 40 fs duration pulse, which has no CEP dependence, are significantly less demanding than with a 10 fs duration pulse. Hence, in what follows, we present results computed for H$_2$ with fixed nuclei driven by a 40 fs duration pulse.

%\begin{widetext}

%%%%%%%%%%%%%%%%%%%% Table for average final momenta (moving)
\begin{table*}
\caption{\label{tab:average_mom}Average final momentum of electron 1 and of the sum of the final electron  momenta along the direction of light propagation   over all double ionization events and over all NSDI  events with one recollision. The values are expressed in terms of $10^{-3}$ a.u. The \% next to NSDI denotes the fraction of NSDI events out of all double ionization events.  } 
\begin{ruledtabular}
\begin{tabular}{lclclclclclclcl}
  & \multicolumn{2}{c} { \hspace{-0.8cm} $\mathrm{H_2}$ $\parallel$ fixed nuclei} & \multicolumn{2}{c} {  \hspace{-0.8cm} $\mathrm{H_2}$  $\perp$ fixed nuclei } &  \multicolumn{2}{c}{  \hspace{-1.4cm} $\mathrm{H_2}$-like atom } \\
\hline
  $\alpha$ & DI & NSDI (91\%) & DI & NSDI (67\%)  & DI & NSDI (89\%) \\
\hline
$\langle \mathrm{p^1_{y,t_{0}}}\rangle_{\alpha}$ &-18  & -14 & -8 & -11 & -36 &  -35\\
$\langle \mathrm{p^1_{y,t_{f}}}\rangle_{\alpha}$ &-63 & -76 & -4 &-4 & \hspace{0.03cm} 95 & \hspace{0.005cm} 96 \\
$\langle \mathrm{p^1_{y,t_{f}}} + \mathrm{p^2_{yt_{f}}}\rangle_{\alpha}$ & -61  & -78 &  \hspace{0.04cm} 6 & \hspace{0.03cm} 3 &  \hspace{0.03cm} 137 & 140 \hspace{0.035cm}\\
\end{tabular}
\end{ruledtabular}
\end{table*}
%\end{widetext}

\subsubsection{Soft versus hard recollision in NSDI}\label{sec:soft_vs_hard_rec}

\begin{figure}[ht]
\includegraphics[scale=0.35]{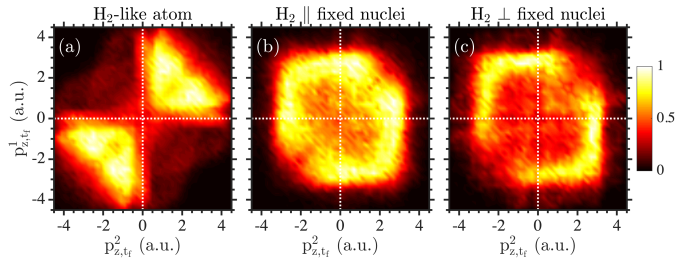}
\caption{Symmetrized correlated electron momenta for NSDI normalized to peak value for (a) the $\mathrm{H_2}-$like atom, (b) $\mathrm{H_2}$ with parallel polarization and fixed nuclei, and (c) $\mathrm{H_2}$ with perpendicular polarization and fixed nuclei.}
\label{fig:correlated_momenta}
\end{figure}

The main difference among the three cases is the relative contribution of the recollision and the laser field in the ionization of electrons 1 and 2. For the H$_2$-like atom, we find that the two electrons  escape following a recollision 
which is assisted by the laser field, i.e. a hard recollision. Specifically, electron 1 returns to the core very close to the initially bound electron. It   transfers energy to electron 2 at times roughly equal to 2T/3 past the time t$_0$. Moreover,
we find that the two electrons ionize at times close to the recollision time. As a result, their final momenta are roughly equal to the value of the vector potential at the time of ionization, i.e. -$\mathrm{\mathbf{A}(y_{i},t_{i}^{ion})}$. Hence, both electrons  escape  with roughly parallel momenta along the direction of polarization, see \fig{fig:correlated_momenta} (a).  

For H$_2$,  we find that the two electrons escape via field ionization assisted by recollision, i.e. soft recollision, mostly so for H$_2$ with perpendicular polarization.  Specifically, we find that electron 1 does not return as close to the core as for a hard recollision. Electron 1 transfers energy to electron 2 at times roughly equal to T/2 past the time t$_0$. Moreover, we find that electron 1 mostly ionizes  shortly after time t$_0$. We note that  electron 1 tunnel-ionizing in the initial state does not imply  ionization at time t$_0$. Indeed, the latter occurs only if the energy of an electron  satisfies   the above mentioned definition of ionization  in terms of the compensated energy. Electron 2 ionizes at times close to the recollision time. The values of the vector potential at times that differ by half a period have opposite signs. Hence,  the electrons escape with opposite momenta along the direction of polarization. This is mostly the case for H$_2$ with perpendicular polarization, see \fig{fig:correlated_momenta}  (c). A more detailed description of soft recollisions is offered in  Ref. \cite{Emmanouilidou3}.   For H$_2$ with parallel polarization, the recollision is not as strong as for the H$_2$-like atom but not as soft as for H$_2$ with perpendicular polarization, see \fig{fig:correlated_momenta}(b).
 This is also corroborated by  NSDI prevailing  in  double ionization  for  H$_2$ with parallel polarization (91\%) and the H$_{2}$-like atom (89\%), see Table \ref{tab:average_mom}. However, for H$_2$ with perpendicular polarization, for a significant ratio of double ionization events  the two electrons escape via field ionization.  The differences between soft and hard recollisions described above  are in accord with our previous findings in the dipole approximation for driven N$_2$ with fixed nuclei and Ar   \cite{Emmanouilidou3,Emmanouilidou4}.
%%%%%%%%%%%%%%%%%%%% Figure: Correlated momenta

The above  differences between soft and hard recollisions result in  the magnetic field jointly with a soft recollision creating a less restrictive gate compared to the one for a hard recollision. This is in accord with our finding, shown  in Table \ref{tab:average_mom}, that the negative shift in the initial transverse momentum of electron 1 is much smaller for  H$_2$ compared to the H$_2$-like atom.

\subsubsection{Electron momentum and position along the  direction  of light propagation  shortly before recollision}\label{sec:electron_momentum_and_position}

Moreover, we show that these different gates  give rise to three key differences in the dynamics of electron 1 in strongly-driven  molecules compared to atoms along the direction of light propagation. Namely, shortly before recollision, electron 1 has mostly small positive momenta for molecules compared to large positive  momenta for atoms.
Also,  shortly before recollision, electron 1 has mostly small positive positions for molecules compared to large negative positions for atoms. Finally, as a result of these two differences,  in the time interval from shortly before recollision  up to the recollision time, the Coulomb forces from the nuclei act to back-scatter electron 1 for molecules and to forward-scatter it  for atoms.

%In what follows, we identify other differences in the  dynamics  of electron 1 when the magnetic field acts together with a soft compared to a hard recollision. 
%First, we investigate  the mapping of the initial transverse momentum of electron 1   from the time of  tunnel-ionization, t$_{0}$, to shortly before recollision, at time $\mathrm{t_{r}-2T/25}$. 

\begin{figure}[ht]
\includegraphics[scale=0.4]{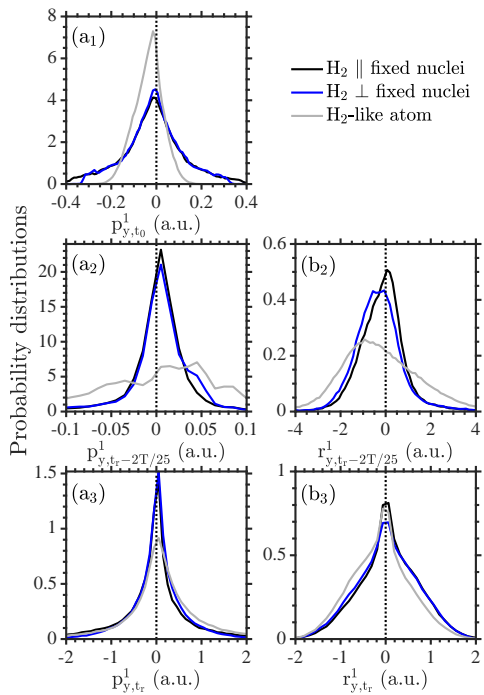}
\caption{Plots of  the distribution of the transverse momentum (a) and position (b)  of electron 1  at the time of tunnel-ionization t$_0$ (a1),  shortly before recollision at time $\mathrm{t_{r}-2T/25}$ ((a2),(b2)), and at the time of recollision $\mathrm{t_{r}}$ ((a3),(b3)).}
\label{fig:distributions}
\end{figure}

\begin{figure*}
\includegraphics[width=\linewidth]{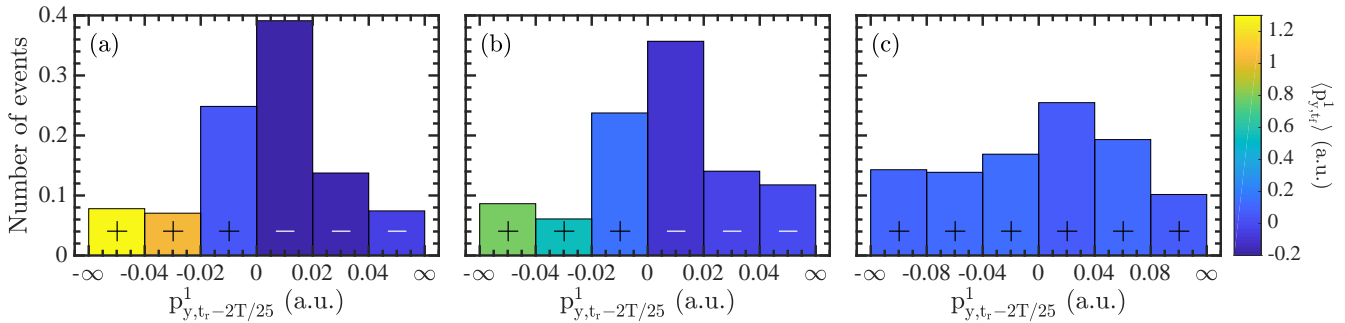}
\caption{Average final  momentum of electron 1 along the y-axis, given by the color scale, as a function of the   momentum of electron 1 along the y-axis  shortly before recollision at time  $\mathrm{t_{r}-2T/25}$ (a) for $\mathrm{H_2}$ with parallel polarization and fixed nuclei, (b) for H$_2$ with perpendicular polarization and fixed nuclei and (c) for the $\mathrm{H_2}$-like atom. The + and - signs are used to facilitate the correct interpretation of the color scale. The height of the columns denote the relative contribution of NSDI events  with momenta at time  $\mathrm{t_{r}-2T/25}$ in a certain momentum interval  to the total number of NSDI events. }
\label{fig:average_momentum_gains}
\end{figure*}

In what follows, we demonstrate in detail these differences between H$_2$ and  the H$_2$-like atom, as well as between the parallel and perpendicular polarization of H$_2$.
In all three cases, the initial transverse momentum of electron 1 is shifted towards negative values at time t$_0$, see \fig{fig:distributions}(a1). 
  Shortly before recollision, we find that  the transverse momentum 
of electron 1 has mostly  large positive values for the H$_2$-like atom and mostly small positive values for  H$_2$, see  \fig{fig:distributions}(a2). 
Moreover, for H$_2$, the transverse momentum of electron 1 has smaller positive values for the parallel compared to the perpendicular polarization, see \fig{fig:distributions}(a2). For H$_2$, it is these small positive values of the transverse electron momentum that give rise  to negative values of the average final  momentum  of electron 1  along the y-axis. This is clearly shown in \fig{fig:average_momentum_gains}(a) and (b). Namely, the electron momenta in the interval [0, 0.02] a.u. shortly before recollision  contribute the most to the average final momentum of electron 1 being negative for H$_2$.
 In contrast,  for the H$_2$-like atom, all electron momenta shortly before recollision contribute to positive values of the average final electron momentum along the y-axis, see   \fig{fig:average_momentum_gains}(c). 
 Also,    comparing the color scale in \fig{fig:average_momentum_gains}(a) with (b) shows  that the  small electron momenta contribute larger negative values to the average final electron momentum  for the parallel compared to the perpendicular polarization.

Next,  shortly before recollision, we investigate the distribution of the position of electron 1 along the direction of light propagation. In all three cases, the position of electron 1 along the y-axis is zero at the exit point at time $\mathrm{t_{0}}$. We find that electron 1 has significantly larger negative values for the  H$_2$-like atom compared to H$_2$, see \fig{fig:distributions}(b2). This difference is in accord with the significantly larger negative shift of the initial transverse momentum of electron 1 for the H$_2$-like atom. Moreover, for H$_2$,   \fig{fig:distributions}(b2) shows that  the distribution of the position along the y-axis of electron 1 peaks at small positive values for parallel polarization. However, it peaks at small negative values for perpendicular polarization. 

\subsubsection{Effect of the Coulomb forces from the nuclei in the time interval from shortly before recollision up to recollision}

Shortly before recollision,  the distributions of the momentum and the position of electron 1 along the direction of light propagation   differ between molecule and atom as well as   between parallel and perpendicular polarization of H$_2$. This results in    the  Coulomb attractive forces exerted on electron 1 from the nuclei having a different effect on the dynamics  of  electron 1. This effect  is more pronounced in the time interval from shortly before recollision up to the time of recollision.

\begin{figure*}
\includegraphics[width=\linewidth]{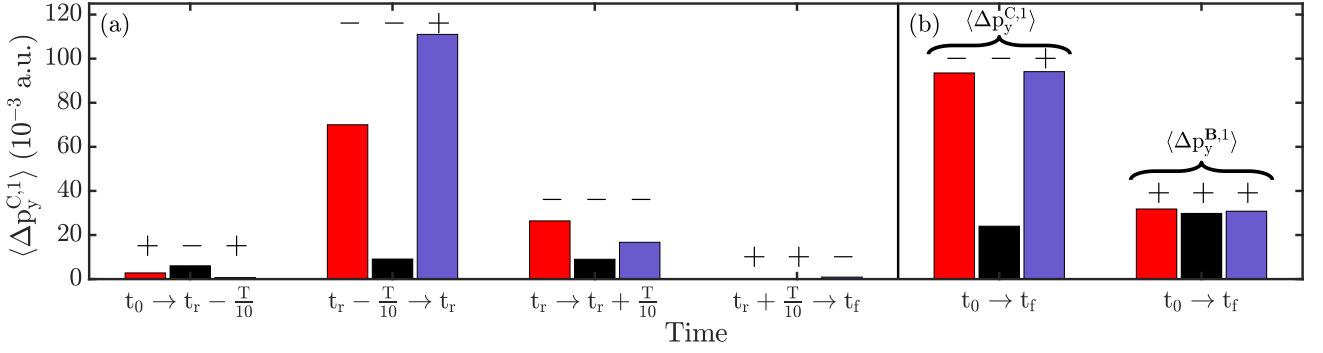}
\caption{Change in momentum of electron 1 due to (a) the Coulomb forces in different time intervals; (b) the total contribution 
to the change in momentum of electron 1 due to  the Coulomb forces and due to the magnetic field.   In both (a) and (b), the three columns in each time interval correspond, from left to right,  to $\mathrm{H_2}$ with parallel polarization and fixed nuclei (red), to  $\mathrm{H_2}$ with perpendicular polarization and fixed nuclei (black), and to the $\mathrm{H_2}$-like atom (purple).  }
\label{fig:changemom}
\end{figure*}

 Indeed, we quantify the contribution of the Coulomb  forces as follows.  The average value of the final electron momentum  along the direction of light propagation  is  equal to
%\begin{align}
%\label{eq:Dp}
%\mathrm{\langle p^1_y,t_f\rangle}&=\mathrm{\langle p^1_y,t_0\rangle+\langle\Delta p^{c,1}_{y,t_0\rightarrow t_b}\rangle+\langle\Delta p^{c,1}_{y,t_b\rightarrow t_r}\rangle} \\
%&+\mathrm{\langle \Delta p^{c,1}_{y,t_r\rightarrow t_a}\rangle+\langle\Delta p^{c,1}_{y,t_a\rightarrow t_f}\rangle+\langle \Delta p^{B,1}_{t_0\rightarrow t_f}\rangle,}
%\end{align}
\begin{align}\label{eq:Dp}
\begin{split}
\mathrm{\langle p^1_y,t_f\rangle}&=\mathrm{\langle p^1_y,t_0\rangle+\langle\Delta p^{C,1}_{y,t_0\rightarrow t_f}\rangle+\langle\Delta p^{B,1}_{y,t_0\rightarrow t_f}\rangle}\\
&=\mathrm{\langle p^1_y,t_0\rangle+\langle\Delta p^{C,1}_{y,t_0\rightarrow t_b}\rangle+\langle\Delta p^{C,1}_{y,t_b\rightarrow t_r}\rangle} \\
&+\mathrm{\langle \Delta p^{C,1}_{y,t_r\rightarrow t_a}\rangle+\langle\Delta p^{C,1}_{y,t_a\rightarrow t_f}\rangle+\langle \Delta p^{B,1}_{t_0\rightarrow t_f}\rangle,}
\end{split}
\end{align}
with
\begin{align}
\begin{split}
\mathrm{\Delta p_{y,t_1\rightarrow t_2}^{C,1}}&=\int_{\mathrm{t_1}}^{\mathrm{t_2}}\left( \dfrac{\mathrm{R}_{\mathrm{y}}^{1}-\mathrm{r}_{\mathrm{y}}^{\mathrm{1}}}{|\mathbf{R}_{1}-\mathbf{r}_{\mathrm{1}}|^3} + \dfrac{\mathrm{R}_{\mathrm{y}}^{2}-\mathrm{r}_{\mathrm{y}}^{\mathrm{1}}}{|\mathbf{R}_{2}-\mathbf{r}_{\mathrm{1}}|^3} + \dfrac{\mathrm{r}_{\mathrm{y}}^{1}-\mathrm{r}_{\mathrm{y}}^{\mathrm{2}}}{|\mathbf{r}_{2}-\mathbf{r}_{\mathrm{1}}|^3} \right)\mathrm{dt}\\
\mathrm{\Delta p_{y,t_1\rightarrow t_2}^{B,1}}&=\int_{\mathrm{t_1}}^{\mathrm{t_2}} \mathrm{F_B} \; \mathrm{dt},
\end{split}
\end{align}
where $\mathrm{\Delta p_{y,t_1\rightarrow t_2}^{C/B,1}}$ denotes the change in momentum of electron 1 in the time interval $\mathrm{[t_1,t_2]}$ due to the Coulomb  forces (C) and  the magnetic field (B). The times $\mathrm{t_b}$ and $\mathrm{t_a}$ correspond to times shortly before $\mathrm{ t_{r}-T/10}$ and shortly after $\mathrm{ t_{r}+T/10}$  recollision. The electric field is always transverse to the direction of light propagation. Hence, it does not contribute to the change in the electron momentum along the y-axis. The Coulomb forces include the attractive forces from  the nuclei and the repulsive one from  electron 2. However,  the largest contribution arises from the Coulomb forces from the nuclei.   The change in electron momentum due to the magnetic field 
   is computed from the start of the propagation $\mathrm{t_0}$ until the final time $\mathrm{t_f}$, i.e. $\mathrm{\langle \Delta p^{B,1}_{t_0\rightarrow t_f}\rangle}$. The reason is that this  change in electron momentum  is roughly the same in all three cases, see 
 \fig{fig:changemom}(b). Also, it is much smaller compared to the change in momentum due to the Coulomb forces  for H$_2$ with parallel polarization and the H$_2$-like atom, see \fig{fig:changemom}(b). In   \fig{fig:changemom}(a), it is  clearly shown   that the change in  momentum of electron 1  due to the Coulomb forces  arises mainly in the time interval from shortly before recollision, $\mathrm{t_{r}-T/10}$, up to the recollision time, $\mathrm{t_{r}}$.

For the H$_2$-like atom,   \fig{fig:changemom}(a) shows that the change in momentum of electron 1 along the y-axis due to the Coulomb forces  is large and positive in the time interval from shortly before recollision and up to the recollision time. This is consistent with our finding in section \ref{sec:electron_momentum_and_position} 
that, shortly before recollision, electron 1 has mostly large negative values of the position  and  mostly large positive momenta along the y-axis. Hence, the Coulomb forces from the nuclei act to mostly accelerate electron 1 along the +y axis in this time interval.

 In contrast, for H$_2$ with parallel polarization, \fig{fig:changemom}(a)  shows that the change in momentum of electron 1 along the y-axis due to the Coulomb forces  is large and negative in the time interval from shortly before recollision and up to the recollision time.
  This is in accord with our finding in section \ref{sec:electron_momentum_and_position} that electron 1 has mostly small positive values of the position and  mostly small positive momenta along the y-axis. As a result,   the Coulomb attractive forces from the nuclei decelerate electron 1 and easily overtake the initially small positive momenta.  This leads to electron 1 eventually accelerating  along the -y-axis and, thus, back-scattering. For H$_2$ with perpendicular polarization,  electron 1 still has mostly small positive momenta shortly before recollision. However, it has mostly small negative positions along the y-axis. As a result, the Coulomb forces act first to accelerate electron 1.  Hence, electron 1 crosses over to the +y axis with larger positive momenta than for the case of parallel polarization. As a result, the Coulomb forces  accelerate electron 1 for a smaller time along the -y-axis.  This results in  the change in electron momentum due to the Coulomb forces being much smaller for H$_2$ with perpendicular compared to parallel polarization, see \fig{fig:changemom}(a).

\subsection{Back-scattering of electron 1 dependence on frequency and intensity for strongly-driven H$_2$ with fixed nuclei}

\begin{figure}[ht]
\includegraphics[scale=0.3]{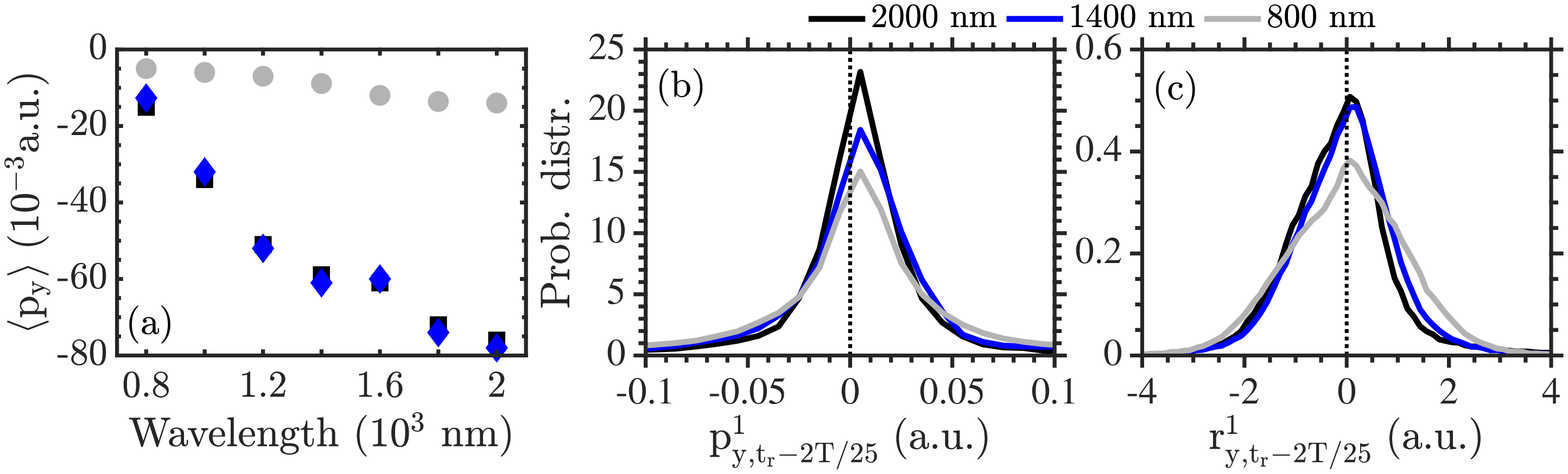}
\caption{In NSDI, for H$_2$ with parallel polarization with fixed nuclei driven by a laser pulse of intensity 3.5$\times$10$^{14}$ Watts/cm$^{2}$, average initial transverse momentum of electron 1 (grey circles), average final transverse momentum of electron 1 (black squares) and average of the sum of the final electron momenta (blue diamonds), as a function of wavelength (a).  Plots of  the distribution of the transverse momentum  (b) and transverse position (c)  of electron 1 shortly before recollision at time $\mathrm{t_{r}-2T/25}$ for three wavelengths.}
\label{fig:wave}
\end{figure}

Next, we demonstrate that back-scattering of electron  1 is a prominent non-dipole effect for a wide range of wavelengths and intensities for two-electron diatomic molecules.
 Indeed, in \fig{fig:wave}(a),   for H$_2$ with parallel polarization and fixed nuclei, we show the dependence on wavelength of the average initial electron momentum, the average final electron momentum  
and the average of the sum of the final electron momenta. Keeping the intensity fixed at 3.5$\times$10$^{14}$ W/cm$^2$, we find that the negative values of all three average electron momenta persist for wavelengths that range from 2000 nm down to 800 nm. However, the negative values of these average electron momenta are large and close to the their values at 2000 nm mainly for wavelengths down to 1400 nm. The  negative values  of these average momenta increase gradually. This  is consistent with the change in the distribution of the momentum (\fig{fig:wave}(b)) and position (\fig{fig:wave}(c)) of electron 1 along the propagation direction  shortly before recollision, at time $\mathrm{t_{r}-2T/25}$. Indeed,  as the wavelength decreases from 2000 nm to   800 nm, we find that electron 1 shortly before recollision has gradually increasing momenta and  increasing positive
positions along the y-axis. These changes  result in  electron 1   forward-scattering  for more NSDI events for smaller than 2000 nm wavelengths.
%%%%%%%%%%%%%%%%%%%% Figure: average as a function of wavelength

\begin{figure}[ht]
\includegraphics[scale=0.3]{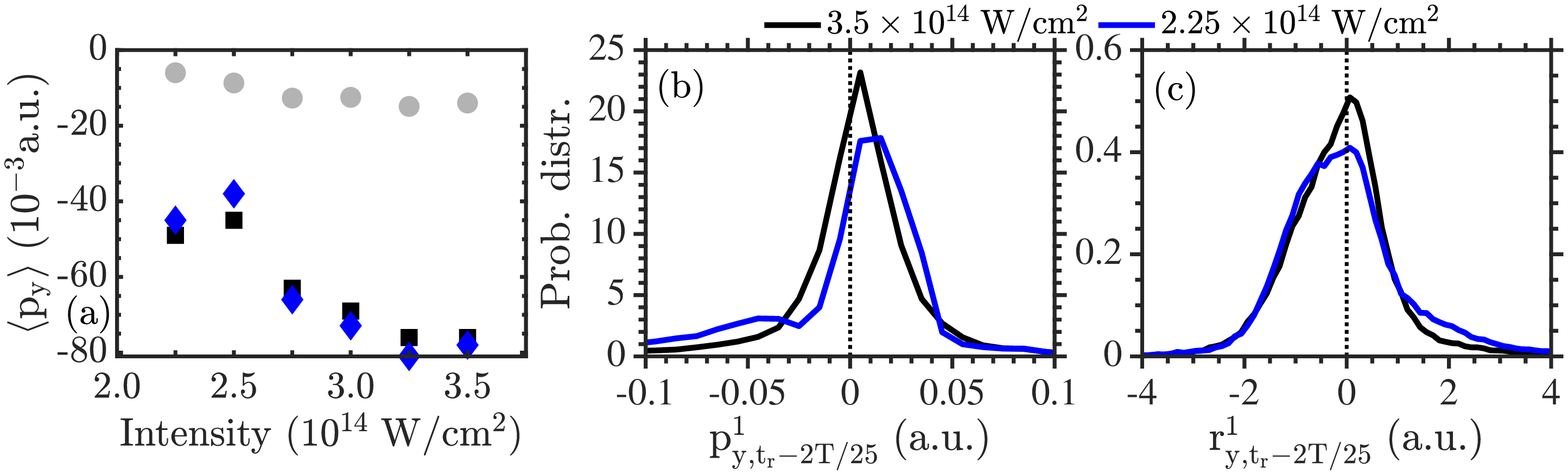}
\caption{In NSDI, for H$_2$ with parallel polarization with fixed nuclei driven  at 2000 nm, average initial transverse momentum of electron 1 (grey circles), average final transverse momentum of electron 1 (black squares) and average of the sum of the final electron momenta (blue diamonds), as a function of intensity (a).  Plots of  the distribution of the transverse momentum  (b) and transverse position (c)  of electron 1 shortly before recollision at time $\mathrm{t_{r}-2T/25}$ for three intensities.}
\label{fig:intensity}
\end{figure}

Also, in \fig{fig:intensity}(a),   for H$_2$ driven at 2000 nm with parallel polarization and fixed nuclei, we show the dependence on intensity of the average initial electron momentum, the average final electron momentum  
and the average of the sum of the final electron momenta. We find that the negative values of all three average electron momenta persist for intensities from 3.5$\times$10$^{14}$ W/cm$^{2}$ down to  2.25$\times$10$^{14}$ W/cm$^{2}$. The  negative values  of these average momenta increase gradually. This  is consistent with the change in the distribution of the momentum (\fig{fig:intensity}(b)) and position (\fig{fig:intensity}(c)) of electron 1 along the propagation direction  shortly before recollision, at time $\mathrm{t_{r}-2T/25}$. As the intensity decreases, we find that electron 1 shortly before recollision has gradually increasing momenta and increasing positive
positions along the y-axis. These changes  result in  electron 1   scattering forwards for more NSDI events for smaller intensities. This dependence on intensity suggests that   back-scattering of electron 1  will be observable even if volume averaging,  i.e. integration over different intensities, is included.

%%%%%%%%%%%%%%%%%%%% Figure: average as a function of intensities

\subsection{Strongly-driven H$_2$ with moving nuclei}

Next, we show that back-scattering of electron 1 along the propagation direction is also present during the Coulomb explosion of the nuclei of strongly-driven H$_2$ with parallel polarization. As discussed above, during the break-up of two-electron diatomic molecules non-sequential double ionization competes with enhanced 
ionization. Non-dipole gated double ionization is not present in enhanced ionization. Hence, in order to increase the contribution of NSDI compared to enhanced ionization we consider a short duration laser pulse. That is, H$_2$ is driven by a near-single-cycle laser field of intensity 3.5$\times$10$^{14}$ W/cm$^2$
 and 10 fs duration  at 2000 nm.  We consider five values of  $\phi$ in Eq. (1), i.e. of CEP, from 0$^{\circ}$ to 360$^{\circ}$ in steps of 60$^{\circ}$. The results presented below are obtained by averaging over the five values of $\phi$.    
\begin{figure}[ht]
\includegraphics[scale=0.4]{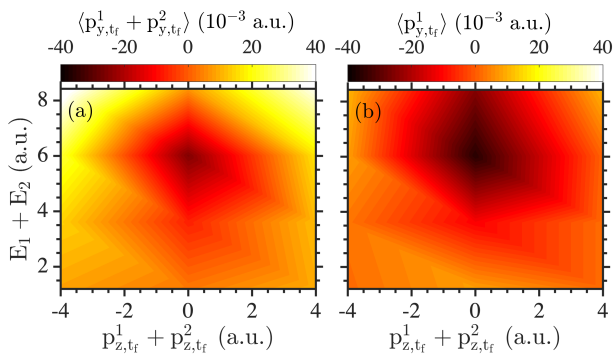}
\caption{Average of the sum of the final electron momenta  along the propagation direction (color scale) (a) and average of the final momentum of electron 1 (color scale) (b) as a function of the sum of the final electron energies  and of the sum of the final electron momenta along the direction of light polarization. }
\label{fig:diff}
\end{figure}

For these laser pulse parameters, we find that enhanced ionization events are more than half of all  double ionization events. Hence, over all double ionization events, it is unlikely that negative values of the average  sum of the final electron momenta will be observed experimentally.  However, we show  that such negative values can be observed when considering doubly differential probabilities. Indeed, as discussed in section \ref{sec:soft_vs_hard_rec}, a  signature of soft recollisions is that the two electrons escape opposite to each other along the direction of polarization, see \fig{fig:correlated_momenta} (b) and (c).
 For these events, the two electrons escape with antiparallel but large  momenta. Hence, for the events that favor electron 1 back-scattering,   $\mathrm{p_{z,t_f}^{1}+p_{z,t_f}^2}\approx 0$ and the sum of the final electron energies $\epsilon_{1}+\epsilon_{2}$ will have values much larger than zero. 
Indeed, in \fig{fig:diff}(a), we show that when both  $\mathrm{p_{z,t_f}^{1}+p_{z,t_f}^2}$ is close to zero and $\epsilon_{1}+\epsilon_2$ is roughly equal to  6 a.u., the average sum of the final electron momenta along the y-axis is roughly equal to -25$\times$10$^{-3}$ a.u.  
 This negative value is due to electron 1 back-scattering. Indeed,  for the same conditions, \fig{fig:diff}(b) shows that electron 1 has a negative average final electron momentum roughly equal to -35$\times$10$^{-3}$ a.u. %%%%%%%%%%%%%%%%%%%% Figure: average as a function of intensities

\section{Conclusions}
We identify a striking signature of non-dipole effects in non-sequential double   ionization of strongly-driven two-electron diatomic molecules. Namely, the average final momentum of the recolliding electron along the direction of light propagation   has a large negative value. As a result, the average sum of the final electron momenta along the direction of light propagation is also large and negative. This large negative values are in contrast to the large positive values for the same quantities  in double ionization of strongly-driven atoms. Hence, the recolliding electron  back-scatters for strongly-driven two-electron diatomic molecules while it forward scatters for strongly-driven atoms. We find this to be the case for intensities much smaller than expected for magnetic fields to arise. Also, we find that back-scattering of the recolliding electron  persists for a wide range of frequencies and intensities. 

The intensities where both magnetic fields effects and recollisions are present correspond to soft recollisions for H$_2$ and hard recollisions for two-electron atoms.
Back versus forward-scattering of the recolliding electron stems from the magnetic field creating a different gate jointly with a soft recollision for molecules compared to a hard recollision for atoms. Both gates result in the initial momentum of the recolliding electron  being mostly negative along the direction of light propagation.
However, this negative shift is smaller for a soft recollision compared to a hard one.  
  These different gates result in three main differences shortly before recollision between molecules and atoms along the direction of light propagation. The recolliding electron    has mostly much smaller  positive momenta for molecules compared to atoms. Also, it has  mostly small positive positions for molecules versus large negative positions for atoms. 
  As a result of the above two differences,  the Coulomb forces from the nuclei act to back-scatter the recolliding electron  in two-electron molecules versus forward-scatter it in atoms. 
  
 Back-scattering of the recolliding electron gives rise to large negative values of the average   sum of the final electron momenta in driven two-electron diatomic molecules. This was demonstrated in the context of H$_2$  with fixed nuclei driven by a pulse with parallel polarization at 2000 nm. Our computations serve as benchmark  for molecules which doubly ionize via recollisions and Coulomb explode much slower than  the duration of the laser pulse. To re-enforce the applicability of our computations for molecules with fixed nuclei, we also consider the break-up of  H$_2$ driven by near-single-cycle laser pulses at 2000 nm. We show that  signatures of back-scattering of the recollinding electron are still found  when considering doubly differential probabilities of double ionization.

%%%%%%%%%%%%%%%%%%%% Table for average final momenta (moving)
%\begin{table*}
%\caption{\label{tab:average_mom_moving}Average momentum of electron 1 and of the sum of the two electrons for the cases under consideration. All the values given are in $10^{-3}$ a.u. } 
%\begin{ruledtabular}
%\begin{tabular}{lclclclclclclcl}
% & \multicolumn{3}{c} {$\parallelsum$ molecule moving nuclei $\langle \Delta p_y \rangle=3$ } & \multicolumn{3}{c} {$\perp$ molecule moving nuclei $\langle \Delta p_y \rangle=8$}\\
%\hline
%  & Total & EI  (66.8\%)  & NSDI  (31.1\%)  & Total & EI (46.3\%)  & NSDI (53.7\%) \\
%\hline
%$\langle \mathrm{p^1_{y,t_{0}}} \rangle$ & -1.1 & 6.3 & -29 & -4  & -4 & -5  \\
%$\langle \mathrm{p^1_{y,t_{f}}} \rangle$ & -1.3 & 6.6 & -16 & 3 & 2 & 3  \\
%$\langle \mathrm{p^1_{y,t_{f}}} + \mathrm{p^2_{yt_{f}}}\rangle$ & 7.4 & 15 & -11 & 6 & 8 & 4 \\
%\end{tabular}
%\end{ruledtabular}
%\end{table*}

   \section*{acknowledgements}
     
A. E. acknowledges  the EPSRC grant no. N031326 and the use of the UCL Myriad High Throughput Computing Facility (Myriad@UCL), and associated support services, in the completion of this work.

\appendix*
\section{Leapfrog Algorithm}\label{Appendix}
In what follows, we describe the leapfrog algorithm.  First, we  initialise the auxiliary variables, $\mathbf{W_{\mathrm{0}}^q}=\mathbf{q}_0,\mathbf{W^{\boldsymbol{\rho}}_{\mathrm{0}}}=\boldsymbol{\rho}_0,\mathrm{W_0^t}=\mathrm{t_0}.$ Then, we propagate for a  time step equal to $\mathrm{h}$, by propagating for half a step each triplet of variables ($\mathbf{q},\mathbf{W}^{\boldsymbol{\rho}}$,t) and ($\mathbf{W^q},\boldsymbol{\rho},\mathrm{W^t}$)  in an alternating way as follows
\begin{align}\label{eq:leapfrog_method}
\begin{split}
% 1st set 
\mathbf{q}_{1/2}&=\mathbf{q}_0+\dfrac{\mathrm{h}}{2}\dfrac{\dot{\mathbf{q}}(\mathbf{W_{\mathrm{0}}^q},\boldsymbol{\rho}_0,\mathrm{W_0^t})}{\Omega(\mathbf{W_{\mathrm{0}}^q})}\\
\mathbf{W}^{\boldsymbol{\rho}}_{1/2}&=\mathbf{W}^{\boldsymbol{\rho}}_{0}+\dfrac{\mathrm{h}}{2}\dfrac{\dot{\boldsymbol{\rho}}(\mathbf{W_{\mathrm{0}}^q},\boldsymbol{\rho}_0,\mathrm{W_0^t})}{\Omega(\mathbf{W_{\mathrm{0}}^q})}\\
\mathrm{t}_{1/2}&=\mathrm{t}_{0}+\dfrac{\mathrm{h}}{2}\dfrac{1}{\Omega(\mathbf{W_{\mathrm{0}}^q})}\\
%2nd set
\mathbf{W_{\mathrm{1}}^q}&=\mathbf{W_{\mathrm{0}}^q}+\mathrm{h}\dfrac{\dot{\mathbf{q}}(\mathbf{q}_{1/2},\mathbf{W}^{\boldsymbol{\rho}}_{\mathrm{1/2}},\mathrm{t}_{1/2})}{\Omega(\mathbf{q}_{1/2})}\\
\boldsymbol{\rho}_{1}&=\boldsymbol{\rho}_{0}+\mathrm{h}\dfrac{\dot{\boldsymbol{\rho}}(\mathbf{q}_{1/2},\mathbf{W}^{\boldsymbol{\rho}}_{1/2},\mathrm{t}_{1/2})}{\Omega(\mathbf{q}_{1/2})}\\
\mathrm{W_1^t}&=\mathrm{W_0^t}+\mathrm{h}\dfrac{1}{\Omega(\mathbf{q}_{1/2})}\\
\mathbf{q}_{1}&=\mathbf{q}_{1/2}+\dfrac{\mathrm{h}}{2}\dfrac{\dot{\mathbf{q}}(\mathbf{W_{\mathrm{1}}^q},\boldsymbol{\rho}_1,\mathrm{W_1^t})}{\Omega(\mathbf{W_{\mathrm{1}}^q})}\\
\mathbf{W}^{\boldsymbol{\rho}}_{1}&=\mathbf{W}^{\boldsymbol{\rho}}_{1/2}+\dfrac{\mathrm{h}}{2}\dfrac{\dot{\boldsymbol{\rho}}(\mathbf{W_{\mathrm{1}}^q},\boldsymbol{\rho}_{1},\mathrm{W_1^t})}{\Omega(\mathbf{W_{\mathrm{1}}^q})}\\
\mathrm{t}_{1}&=\mathrm{t}_{1/2}+\dfrac{\mathrm{h}}{2}\dfrac{1}{\Omega(\mathbf{W_{\mathrm{1}}^q})}\\
\end{split}
\end{align}
The subscripts 0,1/2,1 denote the value of each variable at the initial, half and full time step.

Next, we express  the algorithm that incorporates the leapfrog method in the Bulirsch-Stoer extrapolation scheme over a step H, which  is split into n sub steps of size $\mathrm{h=H/n}$, as follows

\begin{align*}
% 1st set 
\mathbf{q}_{1/2}&=\mathbf{q}_0+\dfrac{\mathrm{h}}{2}\dfrac{\dot{\mathbf{q}}(\mathbf{W_{\mathrm{0}}^q},\boldsymbol{\rho}_0,\mathrm{W_0^t})}{\Omega(\mathbf{W_{\mathrm{0}}^q})}\\
\mathbf{W}^{\boldsymbol{\rho}}_{1/2}&=\mathbf{W}^{\boldsymbol{\rho}}_{0}+\dfrac{\mathrm{h}}{2}\dfrac{\dot{\boldsymbol{\rho}}(\mathbf{W_{\mathrm{0}}^q},\boldsymbol{\rho}_0,\mathrm{W_0^t})}{\Omega(\mathbf{W_{\mathrm{0}}^q})}\\
\mathrm{t}_{1/2}&=\mathrm{t}_{0}+\dfrac{\mathrm{h}}{2}\dfrac{1}{\Omega(\mathbf{W_{\mathrm{0}}^q})}\\
%2nd set
\mathbf{W_{\mathrm{1}}^q}&=\mathbf{W_{\mathrm{0}}^q}+\mathrm{h}\dfrac{\dot{\mathbf{q}}(\mathbf{q}_{1/2},\mathbf{W}^{\boldsymbol{\rho}}_{\mathrm{1/2}},\mathrm{t}_{1/2})}{\Omega(\mathbf{q}_{1/2})}\\
\boldsymbol{\rho}_{1}&=\boldsymbol{\rho}_{0}+\mathrm{h}\dfrac{\dot{\boldsymbol{\rho}}(\mathbf{q}_{1/2},\mathbf{W}^{\boldsymbol{\rho}}_{1/2},\mathrm{t}_{1/2})}{\Omega(\mathbf{q}_{1/2})}\\
\mathrm{W_1^t}&=\mathrm{W_0^t}+\mathrm{h}\dfrac{1}{\Omega(\mathbf{q}_{1/2})}\\
\vdots\\
\mathbf{q}_{\mathrm{m}-1/2}&=\mathbf{q}_{\mathrm{m}-3/2}+\mathrm{h}\dfrac{\dot{\mathbf{q}}(\mathbf{W_{\mathrm{m-1}}^q},\boldsymbol{\rho}_{\mathrm{m}-1},\mathrm{W^t_{\mathrm{m}-1}})}{\Omega(\mathbf{W_{\mathrm{m-1}}^q)}}\\
\mathbf{W_{\mathrm{m-1/2}}^{\boldsymbol{\rho}}}&=\mathbf{W_{\mathrm{m-3/2}}^{\boldsymbol{\rho}}}+\mathrm{h}\dfrac{\dot{\boldsymbol{\rho}}(\mathbf{W^q_{\mathrm{m-1}}},\boldsymbol{\rho}_{\mathrm{m}-1},\mathrm{W^t_{\mathrm{m-1}}})}{\Omega(\mathbf{W_{\mathrm{m-1}}^q)}}\\
\mathrm{t}_{\mathrm{m}-1/2}&=\mathrm{t}_{\mathrm{m}-3/2}+\mathrm{h}\dfrac{1}{\Omega(\mathbf{W_{\mathrm{m-1}}^q)}}\\
\end{align*}
\begin{align*}
% mth set
\mathbf{W^q_\mathrm{m}}&=\mathbf{W^q}_{\mathrm{m}-1}+\mathrm{h}\dfrac{\dot{\mathbf{q}}(\mathbf{q}_{\mathrm{m}-1/2},\mathbf{W_{\mathrm{m-1/2}}^{\boldsymbol{\rho}}},\mathrm{t}_{\mathrm{m}-1/2})}{\Omega(\mathbf{q}_{\mathrm{m}-1/2})}\\
\boldsymbol{\rho}_{\mathrm{m}}&=\boldsymbol{\rho}_{\mathrm{m}-1}+\mathrm{h}\dfrac{\dot{\boldsymbol{\rho}}(\mathbf{q}_{\mathrm{m}-1/2},\mathbf{W_{\mathrm{m-1/2}}^{\boldsymbol{\rho}}},\mathrm{t}_{\mathrm{m}-1/2})}{\Omega(\mathbf{q}_{\mathrm{m}-1/2})}\\
\mathrm{W_\mathrm{m}^t}&=\mathrm{W_{\mathrm{m}-1}^t}+\mathrm{h}\dfrac{1}{\Omega(\mathbf{q}_{\mathrm{m}-1/2})}\\
\vdots\\
\mathbf{q}_{\mathrm{n}}&=\mathbf{q}_{\mathrm{n}-1/2}+\dfrac{\mathrm{h}}{2}\dfrac{\dot{\mathbf{q}}(\mathbf{W^q_\mathrm{n}},\boldsymbol{\rho}_{\mathrm{n}},\mathrm{W^t_\mathrm{n}})}{\Omega(\mathbf{W^q_\mathrm{n}})}\\
\mathbf{W}^{\boldsymbol{\rho}}_{\mathrm{n}}&=\mathbf{W}^{\boldsymbol{\rho}}_{\mathrm{n-1/2}}+\dfrac{\mathrm{h}}{2}\dfrac{\dot{\boldsymbol{\rho}}(\mathbf{W_{\mathrm{n}}^q},\boldsymbol{\rho}_{\mathrm{n}},\mathrm{W_{\mathrm{n}}^t})}{\Omega(\mathbf{W^q_\mathrm{n}})}\\
\mathrm{t}_{\mathrm{n}}&=\mathrm{t}_{\mathrm{n}-1/2}+\mathrm{h}\dfrac{1}{\Omega(\mathbf{W^q_\mathrm{n}})}\\
\end{align*}
where m=2,...,n. 
%Note that when n=1, i.e. h=H, we retrieve the algorithm for the leapfrog method in \eq{eq:leapfrog_method}. 
%%%%%%%%%%%%%%%%%%%%%%%%%%%%%%%%%%%%%%%%%%%%%%%%%%%%%%%%%%
%%%%%%%%%%%%     BIBLIOGRAPHY
%%%%%%%%%%%%%%%%%%%%%%%%%%%%%%%%%%%%%%%%%%%%%%%%%%%%%%%%%%
\bibliography{MagneticfieldBackscattering_bibliography}{}

\end{document}